\documentclass[12pt]{iopart}

\usepackage{iopams}

\usepackage[utf8]{inputenc}     
\usepackage[T1]{fontenc}        
\usepackage{mathrsfs}           
\usepackage{bm}                 
\usepackage{subfig}             
\usepackage[inline]{enumitem}   
\usepackage[colorlinks,
    allcolors=blue]{hyperref}   

\usepackage[inkscapeformat=png]{svg} 
\usepackage{amsopn}             
\usepackage{tikz}               
\usetikzlibrary{decorations.pathmorphing, matrix} 

\usepackage{cite}
\usepackage[]{multirow}
\newcommand\figref[1]{Fig.\,\ref{#1}}
\DeclareMathOperator*{\argmax}{arg\,max}

\begin{document}

\title[]{Cluster Shift Keying: Covert Transmission of Information via Cluster Synchronization in Chaotic Networks}

\author{Zekeriya Sarı$^{1,*}$ and Serkan Günel$^2$}

\address{
    $^1$ The Graduate School of Natural and Applied Sciences, Dokuz Eylül University, İzmir, Turkey}
\address{
    $^2$ Department of Electrical and Electronics Engginering, Dokuz Eylül University, İzmir, Turkey}
\address{
    $^*$ Author to whom any correspondence should be addressed.}
\ead{zekeriya.sari@deu.edu.tr}
\vspace{10pt}
\begin{indented}
    \item[]December 2023
\end{indented}

\begin{abstract}
    A network of chaotic systems can be designed in such a way that the cluster patterns formed by synchronous nodes can be controlled through the coupling parameters. We present a novel approach to exploit such a network for covert communication purposes, where controlled clusters encode the symbols spatio-temporally. The cluster synchronization network is divided into two subnetworks as transmitter and receiver. First, we specifically design the network whose controlled parameters reside in the transmitter. Second, we ensure that the nodes of the links connecting the transmitter and receiver are not in the same clusters for all the control parameters. The former condition ensures that the control parameters changed at the transmitter change the whole clustering scheme. The second condition enforces the transmitted signals are always continuous and chaotic. Hence, the transmitted signals are not modulated by the information directly, but distributed over the links connecting the subnetworks. The information cannot be deciphered by eavesdropping on the channel links without knowing the network topology. The performance has been assessed by extensive simulations of bit error rates under noisy channel conditions.
\end{abstract}

\vspace{2pc}

\section{Introduction}\label{sec:introduction}

The chaotic dynamical systems are characterized by their rich and complex dynamical behavior due to their sensitive dependency on initial conditions and parameter changes, despite their intrinsically simple and deterministic structures \cite{Wiggins2003}. They have been proposed for data encryption \cite{Dachselt1997,Dachselt2001,Kocarev2001,Amigo2009} and secure communication \cite{Bohme1986,Parlitz1992,Kennedy1993,Kolumban1996,feldmann1996communication,Chua1997a} applications due to their high bandwidth and the unpredictable outputs in the long run\cite{Abarbanel1996}.

The networks of coupled dynamical systems can synchronize to follow a single trajectory independent of initial conditions in the case of full synchronization\cite{Pecora1997,Pecora1998,Belykh2006,Chen2007,Yu2009,Aminzare2014,Song2010,Yu2022,Zhao2022,Chen2022}. In a more curious arrangement called \emph{cluster synchronization}, they can synchronize in \emph{clusters} where the nodes are synchronized within the groups, but there is no synchronization among the groups. Numerous studies have shown that a network of continuous-time chaotic systems with a predetermined topology can be forced to exhibit cluster synchronization if the coupling strengths between the nodes in the network exceed certain thresholds. Depending on the network topology, the number of nodes and the coupling type, the required thresholds that make cluster synchronization possible can be determined using the techniques available in Lyapunov stability theory, graph theory, and control theory. Arbitrarily selected cluster patterns in diffusively-coupled undirected networks can be realized by using coupling schemes with cooperative and competetive weights \cite{Ma2006}, pinning control strategy \cite{Wu2009}, graph theory \cite{Schaub2016} or topological sensitivity\cite{Fu2013,Lin2016,Lin2016a,Gambuzza2021}. A desired cluster pattern can also be realized in heterogenous\cite{Wang2019,Lu2010}, adaptive\cite{Wu2009,Wu2011}, fractional-order\cite{Chen2012}, directed\cite{Liu2011}, multiplex\cite{Sevilla2016, Jalan2014a}, and delayed\cite{liu2019cluster,Singh2017} networks. In \cite{Pecora2014}, framework that shows connection between network symmetries and cluster formation is presented. In \cite{Sorrentino2016}, a method is proposed to find and analyze all of the possible cluster synchronization patterns. In \cite{Khanra2022}, a method to identify the clusters of a network without the need of exploring the its symmetry group is introduced. 

In addition, the problem of designing networks that enables the stable realizations of \emph{desired} cluster synchronization with a predefined pattern has been studied extensively. It is also possible to ensure the synchronization of several cluster patterns in the same network consisting of chaotic systems by adjusting the network parameters \cite{hillier2006learning,Belykh2008,belykh2003cluster,Fu2013,Fu2014, Wang2019,Lin2016,Lin2016a}. It is now clear that we can control which clusters of nodes synchronize by changing the coupling strengths between the nodes or gains of the input signals acting on particular nodes \cite{Wang2020,Fu2013}.

The chaotic communication methods are based on the synchronization of chaotic systems in the receiver and the transmitter, and they can be grouped around chaotic masking, chaos shift keying, or bifurcation parameter modulation techniques, generally \cite{Cuomo1993,Kennedy1993,Xiang-Jun2006,Yang2017,Tan2018,Kaddoum2018,Cai2019,Zhang2019,Tan2019,Lu2020,Cai2021}. In the chaotic masking method, the signal to be transmitted is directly added to a chaotic carrier signal. The direct addition can make the systems in the receiver and the transmitter lose their chaotic behaviors unless the power of the message signal is sufficiently small compared to that of the chaotic carrier signal. The chaos shift keying method requires that the outputs of the chaotic systems be switched over a symbol duration according to the message signal and that the switching time is long enough to ensure the synchronization of the chaotic systems in the receiver and the transmitter. This requirement makes its effectiveness controversial in modern applications that require high data transmission rates. In the bifurcation parameter modulation method, the chaotic system parameters are changed by the message signal. However, the sensitive dependence on initial conditions reduces noise immunity. Furthermore, the analyses with time-frequency domain and parameter estimation techniques have shown that these chaotic communication methods cannot prevent the retrieval of the message signal by third parties alone\cite{Short1996,Short1994,Yang1998,Li2003,Li2005,Alvarez2004}. It is hard to admit, but the performances of these once-promising methods are not comparable to those of the classical modulation methods generally, despite their elegant formulation of the problem and implementation simplicity. Arguably, the basic common weakness of the chaos communication systems is their way of modulating the information to change the chaotic dynamics of the systems, directly. In this paper, we propose a novel approach, namely cluster shift keying (CLSK), in which we encode the symbols into the \emph{spatio-temporal synchronization status of a network of chaotic dynamical systems} to mitigate the drawbacks. 
\begin{figure}[!]
    \centering
    \subfloat[The cluster pattern $\mathscr{C}_i$]{
        \includegraphics[width=0.75\textwidth]{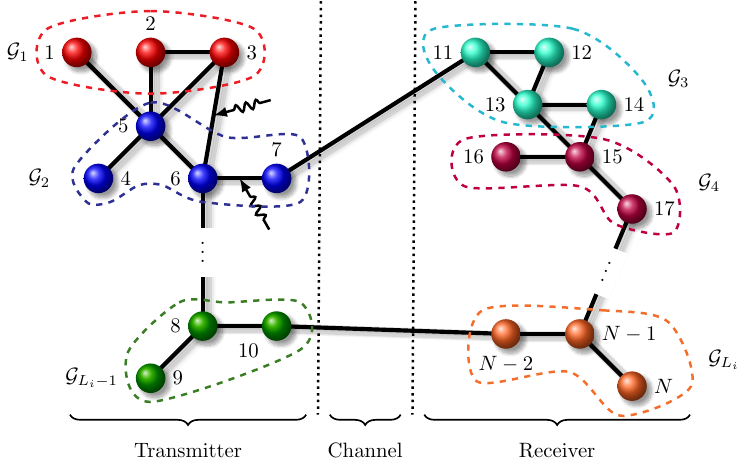}
        \label{subfig: symbol_i}
    }\\
    \subfloat[The cluster pattern $\mathscr{C}_j$]{
        \includegraphics[width=0.75\textwidth]{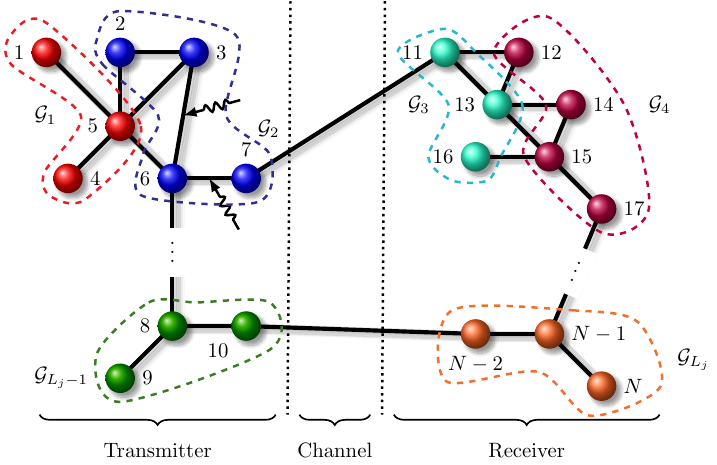}
        \label{subfig: symbol_j}
    }
    \caption{Spatio-temporal status of the network when two different symbols \protect\subref{subfig: symbol_i} $s_i$ and \protect\subref{subfig: symbol_j} $s_j$ encoded with the cluster patterns $\mathscr{C}_i$ and $\mathscr{C}_j$ respectively. The couplings that have the symbol
        (\mbox{\protect\tikz[baseline=-0.5ex]{\protect\draw[>=latex, decorate, decoration = { snake, amplitude = 2pt, segment length = 5pt, pre length = 0.5em, post length = 0.75em}, -latex,] (0, 0) -- ++(0:0.85);}})
        on them are controlled.}
    \label{fig: symbol transmission}
\end{figure}

In our approach, the network is designed specifically to be in several cluster synchronization modes depending on coupling parameters. The cluster synchronization of the network is divided into two subnetworks as transmitter and receiver, and control of particular coupling strengths between the nodes at the transmitter subnetwork switches the synchronization pattern of the structure. The transmitter and receiver subnetworks are not necessarily in drive-response mode \cite{Parlitz1992,Kennedy1993}. The presence of a particular symbol is detected by the presence of the particular cluster pattern at the receiver subnetwork (\figref{fig: symbol transmission}). We choose the cluster patterns such that the nodes connecting the transmitter and receiver are never in the same cluster, essentially. Hence, the signals transmitted always remain chaotic regardless of the symbols being sent, masking the information, effectively. The symbols are represented by the error signals between the nodes at the receiver subnetworks. The detection of a symbol is done over the nodes at the receiver, and the channel links are not directly involved. As a result, to decipher the symbols one needs complete knowledge of the network topology, rendering the communication immune to eavesdropping. Besides, the scheme tends to be robust against channel noise due to the guaranteed asymptotic stability of the network by construction.

In the sequel, we revisit the conditions to design networks capable of achieving various cluster synchronization patterns (Section \ref{sec: network design}), first. Section \ref{sec: communication system design} presents the conceptual design of the proposed communication system based on such networks. The assessment of the performance of the proposed scheme is given in Section \ref{sec: numerical results} through a proof of concept design.

\section{Cluster Synchronization}
\label{sec: network design}

Consider a network of $N$ identical linearly-coupled continuous-time chaotic systems whose dynamics evolves by
\begin{equation}
    \dot{\bm{x}}_i = f(\bm{x}_i) + \epsilon \sum_{j = 1}^N \xi_{ij} \bm{\Gamma} \bm{x}_j,  \quad i = 1, 2, \ldots, N,
    \label{eq:-odenet}
\end{equation}
where  $\bm{x}_i \in \mathbb{R}^d$ is the state vector of $i$-th node, and $f$ is the function defining the individual node dynamics. $\epsilon > 0$ is the coupling strength.
$\bm{\Xi} = [\xi_{ij}] \in \mathbb{R}^{N \times N}$ determines the network topology: if the nodes $i$ and $j$ are coupled to each other, $\xi_{ij} > 0$; otherwise $\xi_{ij} = 0$.
The inner coupling matrix $\bm{\Gamma}$ determines by which state variables the nodes are coupled.
We assume that the network is diffusively-coupled, hence, $\bm{\Xi}$ is a zero-row-sum matrix, i.e., $\sum_{j = 1}^N \xi_{ij} = 0, \; \forall i$, and the eigenvalues of $\bm{\Xi}$ satisfy $0 = \lambda_1 > \lambda_2 \geq \lambda_3 \geq \ldots \geq \lambda_N$.
Associated with the topology of the network in \eref{eq:-odenet}, an undirected graph $\mathscr{G} = \{\mathcal{N}, \mathcal{L}\}$ can be constructed, where $\mathcal{N} = \{1, 2, \ldots, N\}$ is the set of nodes, and $\mathcal{L} = \{\ell_{ij} = (i, j) ~|~ i, j \in \mathcal{N}, i < j \}$ is the set of links $\ell_{ij}$ that connects the nodes $i$ and $j$.

Let $\mathcal{N}$ be partitioned such that $\mathcal{N} = \cup_{l = 1}^L \mathcal{G}_l$ where $\mathcal{G}_l \neq \emptyset, \forall l$, and $\mathcal{G}_i \cap \mathcal{G}_j = \emptyset, i \neq j$, and $\mathcal{G}_l$ represents the node indices in cluster $l$. Let $\bm{e}_{ij}(t) = \bm{x}_i(t) - \bm{x}_j(t)$. The cluster synchronization is achieved if
\numparts
\begin{eqnarray}
    \lim_{t \mapsto \infty} \Vert \bm{e}_{ij}(t) \Vert & = 0, \quad i, j \in \mathcal{G}_k, i \neq j                                \\
    \lim_{t \mapsto \infty} \Vert \bm{e}_{ij}(t) \Vert & \neq 0, \quad i \in \mathcal{G}_k, j \in \mathcal{G}_l, k \neq l, i \neq j
    \label{eq-cluster-synchronization-definition}
\end{eqnarray}
\endnumparts
for $k, l = 1, 2, \ldots, L$ where $\dot{\bm{x}}_i(t) \not\equiv 0, \forall t, \forall i \in \mathcal{N}$. Thus, when the cluster synchronization is achieved, the nodes in the same cluster synchronize while those in different clusters are out of synchronization.

The spatial status of the network corresponding to a cluster synchronization is called a \emph{cluster pattern} represented by $\mathscr{C} = \{\mathcal{G}_1, \mathcal{G}_2, \ldots, \mathcal{G}_L\}$. If certain conditions are satisfied, a network may exhibit several cluster patterns $\mathscr{C}_m = \{\mathcal{G}_{1, m}, \mathcal{G}_{2, m}, \ldots, \mathcal{G}_{L_m, m}\}, m = 1, 2, \ldots, M$ where $L_m$ is the number of clusters in $\mathscr{C}_m$, and $\mathcal{N} = \cup_{l = 1}^{L_m} \mathcal{G}_{l, m}, \; \forall m$. We call $\mathscr{P} = \{\mathscr{C}_1, \mathscr{C}_2, \ldots, \mathscr{C}_M \}$ and $M$ the cluster pattern set and the cluster pattern capacity of the network, respectively.

The number of different cluster patterns that can be observed in a network is primarily determined by the number of topological symmetries of the network \cite{Fu2013}. For a network consisting of $N$ nodes, each topological symmetry $\mathcal{S}$ of the network can be associated with a permutation matrix $\bm{\Delta} = [\delta_{ij}] \in \mathbb{R}^{N \times N}$ where $\delta_{ij} = 1$ if $(i, j)$ is a symmetric pair, i.e., the exchange of nodes $i$ and $j$ with respect to $\mathcal{S}$ does not change the topological structure; otherwise, $\delta_{ij} = 0$. A cluster pattern $\mathscr{C}$ can be constructed by grouping the nodes forming the symmetric pairs into the same clusters.

Although different cluster patterns are obtained for different topological symmetries, the cluster synchronizations corresponding to these cluster patterns may be either dynamically or topologically unstable \cite{Lin2016a}. To make the cluster pattern stable, consider a control network whose dynamics evolves by
\begin{equation}
    \dot{\bm{\tilde{x}}}_k = f(\bm{\tilde{x}}_k) + \sum_{l = 1}^L \omega_{kl} \bm{\Gamma} \bm{\tilde{x}}_l, \quad k = 1, 2, \ldots, L,
    \label{eq-control-network}
\end{equation}
and
\begin{equation}
    \omega_{kl} = \sum_{j \in \mathcal{G}_k} \sum_{j \in \mathcal{G}_l} \xi_{ij} / N_k,
    \label{eq-control-weights}
\end{equation}
with $N_k$ is the number of nodes in $\mathcal{G}_k$. The original network in (\ref{eq:-odenet}) is controlled by pinning each node $\bm{\tilde{x}}_k$ to all nodes of $\mathcal{G}_k$. Thus, the dynamics of the controlled network takes the form
\begin{equation}
    \dot{\bm{x}}_i = f(\bm{x}_i) + \epsilon \sum_{j = 1}^N \xi_{ij} \bm{\Gamma} \bm{x}_j + \alpha \epsilon \sum_{k = 1}^L \delta_{ik} \bm{\Gamma} \bm{\tilde{x}}_k, \quad i = 1, 2, \ldots, N,
    \label{eq-controlled-network}
\end{equation}
where $\alpha > 0$ is the control strength.

Let us assume the control network is deactivate first, i.e. $\alpha=0$, and investigate the stability of the cluster pattern $\mathcal{C}$ with respect to a topological symmetry $\mathcal{S}$. Consider the special solution $\bm{s}_k$ with $\dot{\bm{s}}_k = f(\bm{s}_k)$ associated to cluster $\mathcal{G}_k$ for $k = 1, 2, \ldots, L$. The perturbation $\bm{e}_i = \bm{x}_i - \bm{s}_k$ where $i \in \mathcal{G}_k$ evolves by
\begin{equation}
    \dot{\bm{e}}_i = Df(\bm{s}_k) \bm{e}_i + \epsilon \sum_{j \in \mathcal{G}_k} \xi_{ij} \bm{\Gamma}, \bm{e}_j + \epsilon \sum_{l \neq k} \sum_{m \in \mathcal{G}_l} \xi_{im} \bm{\Gamma}(\bm{e}_m - \bm{e}_i),
    \label{eq-variational-equation}
\end{equation}
where $Df$ is the Jacobian of $f(\cdot)$.

Denoting $\bm{E} = [\bm{E}_1, \bm{E}_2, \ldots, \bm{E}_L]^T$, where $\bm{E}_k = [\bm{e}_1, \bm{e}_2, \ldots , \bm{e}_{N_k}]^T$ is the error vector corresponding to $k$-th cluster, (\ref{eq-variational-equation}) can be written as,
\begin{equation}
    \dot{\bm{E}} = \left[ \sum_{k = 1}^L \bm{P}_k \otimes Df(\bm{s}_k) + \epsilon \Xi \sum_{k = 1}^L \bm{P}_k \otimes \bm{\Gamma} \right] \bm{E},
    \label{eq-varitional-equation-in-compact-form}
\end{equation}
where $\bm{P}_k = [p_{ik}]$, $p_{ik} = 1$ if $i \in \mathcal{G}_k$; $p_{ik} = 0$, otherwise. Let $\bm{\Delta}$ be the permutation matrix corresponding to $\mathcal{S}$. $\bm{\Delta}^2 = \bm{I}_{N\times N}$ by construction, therefore $\bm{\Delta}$ is diagonalizable. Thus, there exist $\bm{\Psi} = [\bm{\psi}_1, \bm{\psi}_2, \ldots, \bm{\psi}_N]^T \in \mathbb{R}^{N \times N}$ such that $diag(\zeta_1, \zeta_2, \ldots, \zeta_N) = \bm{\Psi}^{-1} \bm{\Delta} \bm{\Psi}$ where $\zeta_k$ is the eigenvalue corresponding to the eigenvector $\bm{\psi}_k$ of $\bm{\Delta}$. Using the transformation $\bm{Y} =  h(\bm{E}) = \bm{\Psi} \bm{E}$, (\ref{eq-varitional-equation-in-compact-form}) can be transformed into
\begin{equation}
    \dot{\bm{Y}} = \left[ \sum_{k = 1}^L \bm{P}_k \otimes Df(\bm{s}_k) + \epsilon \bm{\Theta} \sum_{k = }^L \bm{P}_k \otimes \bm{\Gamma}  \right] \bm{Y},
    \label{eq-variational-eqauation-compact-transformed}
\end{equation}
where $\bm{Y} = [\bm{Y}_1, \bm{Y}_2, \ldots, \bm{Y}_L]^T$,
\begin{equation}
    \bm{\Theta}  = \bm{\Psi}^{-1} \bm{\Xi} \bm{\Psi} =
    \left[\matrix{
            \bm{\Omega} & \bm{0}    \cr
            \bm{0} & \bm{\Phi} \cr
        }\right],
    \label{eq-block-diagonal-transformed-matrix}
\end{equation}
with $\bm{\Omega} = diag(\bm{\Omega}_1, \bm{\Omega}_2, \ldots, \bm{\Omega}_L), \bm{\Omega}_k \in \mathbb{R}^{(N_k - 1) \times (N_k - 1)}$ and $\bm{P} \in \mathbb{R}^{L \times L}$.  $\bm{\Omega}_k$ characterizes the motion transverse to the synchronization manifold of $k$-th cluster, and the subspace spanned by the eigenvectors of $\bm{\Omega}_k$ is called transversal subspace of $k$-th cluster. Since $\bm{\Phi}$ characterizes the motion parallel to the synchronnization manifolds of the clusters, the subspace spanned by the eigenvectors of $\bm{\Phi}$ is called synchronuous subspace.

The transformation $h$ decouples the dynamics of the perturbations corresponding to each cluster. Hence, the stability of $k$-th cluster can be determined from the decoupled dynamics
\begin{equation}
    \dot{\bm{Y}}_k = [Df(\bm{s}_k) + \epsilon \bm{\Omega_k} \bm{\Gamma} ] \bm{Y}_k,
    \label{eq-variation-equations-of-kth-cluster}
\end{equation}
where $\bm{Y}_k = [\bm{y}_1, \bm{y}_2, \ldots, \bm{y}_{N_k}]$, and $\bm{I}_k$ is $N_k - 1$ dimensional identity matrix. Let $0 > \lambda_{k, 1}^t \geq \lambda_{k, 2}^t \geq \ldots \geq \lambda_{N_k}^t$ be eigenvalues, and $\bm{\rho}_{k, 1}, \bm{\rho}_{k, 2}, \ldots, \bm{\rho}_{k, N_k}$ be the corresponding eigenvectors of $\bm{\Omega_k}$. Using the transformation $\bm{Z}_k = q(\bm{R}^{-1} \bm{Y}_k)$ where $\bm{R} = [\bm{\rho}_{k, 1}, \bm{\rho}_{k, 2}, \ldots, \bm{\rho}_{k, N_k}]$ and $diag(\lambda_{k, 1}^t, \lambda_{k, 2}^t, \ldots, \lambda_{k, N_k}^t) = \bm{R}^{-1} \bm{\Omega}_k \bm{R}$, the dynamics of $l$-th transverse mode  $\bm{z}_{k,l}$ can be obtained as
\begin{equation}
    \dot{\bm{z}}_{k, l} = [Df(\bm{s}_k) + \epsilon \lambda_{k, l}^t \bm{\Gamma}] \bm{z}_{k, l}, \quad l = 1, 2, \ldots, N_k.
    \label{eq-variation-equations-of-kth-cluster-transformed}
\end{equation}
Dropping the subscripts and superscripts, and letting $\eta = \epsilon \lambda$ in \eref{eq-variation-equations-of-kth-cluster-transformed}, we obtain the generic system
\begin{equation}
    \dot{\bm{z}} = (Df(\bm{s}) + \eta \bm{\Gamma}) \bm{z}.
    \label{eq: linearized system}
\end{equation}
The master stability function $\mu = g(\eta)$ maps $\eta$ to the maximum Lyapunov exponent $\mu$ of the system in \eref{eq: linearized system}. If $\mu_{k,l} = g(\epsilon \lambda_{k, l}^t) \leq 0$, then $l$-th mode is stable, i.e., $\lim_{t \mapsto \infty} \Vert \bm{z}_{k,l} \Vert = 0$; otherwise it is unstable.

Let $0 = \lambda_{1}^s > \lambda_2^s \geq \ldots \geq \lambda_{L}^s$ be eigenvalues of $\bm{\Phi}.$ Then, the following conditions must be necessarily satisfied for a cluster synchronization manifold to be stable \cite{Lin2016a}:
\begin{enumerate}[label=\roman*)]
    \item All the transverse modes in the transverse subspace must be stable, i.e., $\mu_{k,l}^t = g(\epsilon \lambda_{k,l}^t) \leq 0, \; \forall k, l$.
    \item At least one of nontrivial modes in the synchronous subspace be unstable, i.e., $\mu_{k}^s = g(\epsilon \lambda_{k}^s) > 0, \exists k \in \{ 2, 3, \ldots, L\}$ to avoid full synchronization.
\end{enumerate}
There exists a threshold $\tilde{\eta} < 0$ such that $\mu = g(\eta) < 0$ for $\eta < \tilde{\eta}$ for most of the well-known dynamical systems \cite{Huang2009}. Denoting $\lambda_{min} = \min\{|\lambda_{k, 1}^t|, k = 1, 2, \ldots, L\}$, the necessary conditions for the cluster synchronization are satisfied if
\begin{equation}
    |\lambda_{2}^s| < \lambda_{min}
    \label{eq-eigvalue-condition}
\end{equation}
and 
\begin{equation}
    \epsilon \in [\tilde{\eta} / \lambda_{min}, \; \tilde{\eta} / \lambda_{2}^s].
\end{equation}

When the control network is activated, it can be seen that the matrix in (\ref{eq-block-diagonal-transformed-matrix}) takes the form
\begin{equation}
    \bm{\Theta}  = \bm{\Psi}^{-1} \bm{\Xi} \bm{\Psi} =
    \left[\matrix{
            \bm{\Omega} - \alpha I_{N^\prime} & \bm{0}              & \bm{0}\cr
            \bm{0}                                        & \bm{\Phi} - \alpha I_L & \alpha \bm{I}_L \cr
            \bm{0}                                        & \bm{0}              & \bm{A}\cr
        }\right]
    \label{eq-block-diagonal-transformed-matrix-shifted}
\end{equation}
where $\bm{A} = [\alpha_{kl}] \in \mathbb{R}^{L \times L}$ is the control matrix in (\ref{eq-control-network})\cite{Lin2016a} . Correspondingly, the condition for inducing cluster synchronization is $\alpha > \alpha_1 = \bar{\eta} / \epsilon - \lambda_{min}$. In the case of induced cluster synchronization, the nodes in the same cluster in the original network in (\ref{eq-controlled-network}) are synchronized while there is no synchronization between a control node $\hat{\bm{x}}_k$ and its assoicated controlled nodes $\overline{\bm{x}}_i, i \in \mathcal{G}_k$. Hence, the synchronization is induced, not controlled, by the control node. If the control strength $\alpha$ satisfies $\alpha > \alpha_2 = \bar{\eta} / \epsilon$, the control node $\hat{\bm{x}}_k$ and all its associated controlled nodes $\overline{\bm{x}}_i, i \in \mathcal{G}_k$ are synchronized. Thus, the cluster synchronization is controlled by the control node if $\alpha > \alpha_2$.

\section{Communication System Design}
\label{sec: communication system design}

\subsection{Network Design Requirements}
\label{subsec: network design}

We need a programmable network in the proposed scheme. Consider a network as in \eref{eq:-odenet}. $\mathcal{N}$ can be partitioned such that $\mathcal{N} = \mathcal{N}_T \cup \mathcal{N}_R$ where $\mathcal{N}_T = \{i_{1}, i_{2}, \ldots, i_{N_T}\}$ and $\mathcal{N}_R = \{j_{1}, j_{2}, \ldots, j_{N_R}\}$ are the sets of node indices in the transmitter and receiver, respectively.  Let $\mathcal{L}_T =  \{\ell_{ij} ~|~ i, j \in \mathcal{N}_T\} \cap \mathcal{L} $ and $\mathcal{L}_R =  \{\ell_{ij} ~|~ i, j \in \mathcal{N}_R\} \cap \mathcal{L} $  be the sets of links in the transmitter and receiver, respectively. Then, the set of the links in the channel is $\mathcal{L}_C = \mathcal{L} \backslash (\mathcal{L}_T \cup \mathcal{L}_R)$.

The network \eref{eq:-odenet} must satisfy the following additional requirements:
\begin{enumerate}[label=\roman*)]
    \item The network must have a cluster pattern capacity of \mbox{$M>1$}.
    \item It is possible to switch from one cluster synchronization pattern to the other by means of predetermined control parameters, such as the coupling or control strengths,
    \item The nodes can be distributed between the transmitter and receiver such that the nodes of the links in $\mathcal{L}_C$ do not synchronize  throughout the whole symbol transmission, i.e, $\frac{d \Vert \bm{e}_{ij}(t) \Vert }{dt} \not\equiv 0, \forall t, \forall \ell_{ij} \in \mathcal{L}_C$.
    \item The distribution of nodes can be performed in such a way that the parameters that control the cluster patterns of the whole network stay in the transmitter side only. In other words, the control parameters must be associated with elements of $\mathcal{L}_T$.
\end{enumerate}
Besides, we must be able to detect the spatio-temporal status of the receiver subnetwork over $\mathcal{N}_R$ only. The coupling strengths of  some links in $\mathcal{L}_T$ can be chosen as the control parameters, which must alter whole networks' synchronization status by ensuring the existence of different cluster patterns  $\mathscr{C}_i \in \mathscr{P}, i = 1, 2, \ldots, M$.

\figref{fig: symbol transmission} illustrates a network that satisfies these requirements. \figref{subfig: symbol_i} and \ref{subfig: symbol_j} illustrate the cluster patterns $\mathscr{C}_i, \mathscr{C}_j \in \mathscr{P}$ chosen arbitrarily depending on the control links $\ell_{36}$ and $\ell_{67}$.

\subsection{Symbol Encoding}
\label{subsec: symbol encoding}

Provided a programmable network such as in \figref{fig: symbol transmission}, we encode information by mapping symbols $s_i$ to cluster patterns $\mathscr{C}_i \in \mathscr{P}, \forall i$ one to one. During the transmission of $s_i$, the control parameters are adjusted to ensure $\mathscr{C}_i$ in the network. The detection of $s_i$ is performed by detecting the presence of $\mathscr{C}_i$ in the receiver. In \figref{subfig: symbol_i} and \ref{subfig: symbol_j}, the arbitrarily chosen $s_i$ and $s_j$ are encoded with $\mathscr{C}_i$ and $\mathscr{C}_j$, respectively. The nodes of the links in $\mathcal{L}_C$ are never in the same cluster for all symbol transmissions in this \emph{cluster shift keying} scheme. This choice of topology guarantees that the transmitted signals are always chaotic, and not modulated by the symbol being sent directly. Instead, the information is distributed over the whole network. Hence, the transmitted information cannot be extracted from the channel signals without the knowledge of the transmitter and receiver topologies and symbol-to-cluster-pattern mapping even if the channel is eavesdropped.

\subsection{Noise Model}
\label{subsec: noise model}

In the presence of Gaussian distributed noise affecting the links, the coupled network can be represented with a stochastic differential equation (SDE)
\begin{equation}
    d\bm{x}_i = \left( f(\bm{x}_i) + \epsilon \sum_{j = 1}^N \xi_{ij} \bm{\Gamma} \bm{x}_j + \alpha \epsilon \sum_{k = 1}^L \delta_{ik} \bm{\Gamma} \bm{\tilde{x}}_k \right) dt + \alpha \epsilon \sigma \sum_{k = 1}^L \delta_{ik} \bm{\Gamma} d\bm{W}_{k},
    \label{eq: sdenet}
\end{equation}
where $\bm{W}_{k}$ is the Wiener process acting on the link $\ell_{ik}$. $\sigma > 0$ is the noise strength.  Denoting $\bm{x} = [\bm{x}_1^T, \bm{x}_2^T, \ldots, \bm{x}_{N}^T]^T$, $\bm{\tilde{x}} = [\bm{\tilde{x}}_1^T, \bm{\tilde{x}}_2^T, \ldots, \bm{\tilde{x}}_{L}^T]^T$,  $F(\bm{x}) = [f(\bm{x}_1)^T, f(\bm{x}_2)^T, \ldots, f(\bm{x}_{N})^T]^T$, $F(\bm{\tilde{x}}) = [f(\bm{\tilde{x}}_1)^T, f(\bm{\tilde{x}}_2)^T, \ldots, f(\bm{\tilde{x}}_{L})^T]^T$, $d\bm{W} = [d\bm{W}_{1}^T, d\bm{W}_{2}^T, \ldots, d\bm{W}_{L}^T]^T$, the network in \eref{eq: sdenet} can be represented more compactly as
\begin{eqnarray}
    \left[
        \matrix{
            d\bm{\tilde{x}} \cr
            d\bm{x} \cr
        }\right]
    &=
    \{
    \left[
        \matrix{
            F(\bm{\tilde{x}}) \cr
            F(\bm{x}) \cr
        }\right]
    +
    \left(
    \left[
        \matrix{
        \bm{A}                 & \bm{0} \cr
        \alpha \epsilon \bm{\Delta} & \epsilon \bm{\Xi} \cr
        }\right]
    \otimes
    \bm{\Gamma}
    \right)
    \left[
        \matrix{
            \bm{\tilde{x}} \cr
            \bm{x} \cr
        }\right]
    \} dt \\
    &+
    \left(
    \left[
        \matrix{
        \bm{0} \cr
        \sigma \alpha \epsilon \bm{\Delta} \cr
        }\right]
    \otimes \bm{\Gamma} \right) d\bm{W} \nonumber
    \label{eq: sdenet compact},
\end{eqnarray}

The noise in \eref{eq: sdenet} is not additive as it is directly injected into the dynamics of the network. Unfortunately, the non-additive nature of the noise avoids direct comparison of noise performance to conventional communication schemes.

\subsection{Detection Rule}
\label{subsec: detection rule}

Let $T_b \in \mathbb{R}^{+}$ be the symbol duration. We detect $n$-th symbol $s[n] \in \{s_1, s_2, \ldots, s_M\}$, during the time frame \mbox{$nT_b \leq t \leq (n + 1) T_b$}, \mbox{$n = 0, 1, \ldots$} by detecting the corresponding cluster pattern $\mathscr{C}[n] \in \mathscr{P}$ in the receiver.  The energy of the synchronization error between the nodes $i$ and $j \in \mathcal{N}_R$ when transmitting $s[n]$ is
\begin{equation}
    E_{ij}[n] = \int_{n T_b}^{ (n + 1) T_b} \Vert \bm{e}_{ij} (t)  \Vert^2 dt \; .
\end{equation}
We decide that the node pair ($i, j$) is synchronized  if  \mbox{$E_{ij}[n] \leq \gamma_n$} given some threshold $\gamma_n > 0$. Specifically,
\begin{equation}
    \gamma_n = \frac{1}{N_R^2} \sum\limits_{i = 1}^{N_R}\sum\limits_{j = 1}^{N_R} E_{ij}[n] \; .
\end{equation}
We represent $\mathscr{C}[n]$ by a synchronization matrix \mbox{$\bm{A}[n] = [\alpha_{ij}[n]] \in \mathbb{R}^{N_R \times N_R}$} where
\begin{equation}
    \alpha_{ij}[n] =
    \cases{
        1, & $E_{ij}[n] \leq \gamma_n$, \\
        0, & $E_{ij}[n] > \gamma_n$,
    } \quad \forall i, j  \in \mathcal{N}_R.
    \label{eq: synchronization matrix}
\end{equation}
Let $\bm{B}_{m} = [\beta_{m,ij}] \in \mathbb{R}^{N_R \times N_R}, m = 1, 2, \ldots, M$ be the reference matrices corresponding to the cluster patterns $\mathscr{C}_m$, where
\begin{equation}
    \beta_{m, ij} =
    \cases{
        1, & $i, j \in \mathcal{G}_{p,m}$                                           \\
        0, & $i \in \mathcal{G}_{p,m}, \; j \in \mathcal{G}_{q, m}, p \neq q,$
    }
\end{equation}
$p, q = 1, 2, \ldots, L_m$ and $i, j \in \mathcal{N}_R$. Then, we use the detection rule
\begin{equation}
    \hat{s}[n]  = \argmax_{m = 1, 2, \ldots, M} \bigl[ h(\bm{A}[n] \odot \bm{B}_m) \bigr]
    \label{eq: detection rule}
\end{equation}
where $\hat{s}[n]$ is the $n$-th detected symbol, $h(\bm{Z}) = \sum_{i = 1}^N \sum_{j = 1}^N \zeta_{ij}$ for $\bm{Z} = [\zeta_{ij}] \in \mathbb{R}^{N \times N}$, and $\odot$ denotes the Hadamard product of two matrices.

Note that we are not concerned about the exact waveforms of the chaotic signals transmitted over $\mathcal{L}_C$ in detection. In other words, the received signals, which do not carry any directly modulated information, are not used in the detection process essentially. The signals on $\mathcal{L}_C$ provide interaction between the control parameters in the transmitter and the subnetwork in the receiver indirectly. The control parameters are adjusted for the symbol to be sent. The corresponding cluster pattern of the subnetwork in the receiver is checked through the signals over the links in $\mathcal{L}_R$, from which the transmitted information is extracted.

\section{Illustrative Examples}
\label{sec: numerical results}

\subsection{Example I: Design without Control Network}
\label{subsec-an-example-network-without-control}

\subsubsection{The Network}
\begin{figure}
    \centering
    \subfloat[]{
        \includegraphics[width=0.32\textwidth]{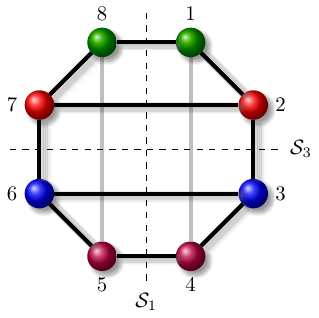}
        \label{subfig: octagon cyclic5}
    }\hfil
    \subfloat[]{
        \includegraphics[width=0.32\textwidth]{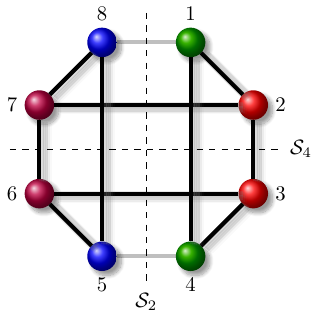}
        \label{subfig: octagon cyclic6}
    } \\
    \subfloat[]{
        \includegraphics[width=0.42\textwidth]{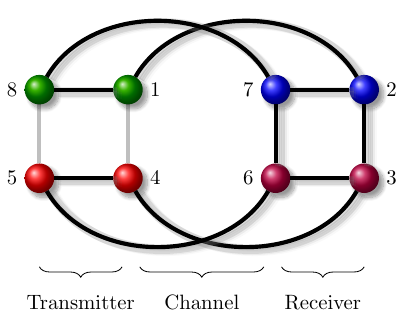}
        \label{subfig:  octagon cyclic5 divided}
    }\hfil
    \subfloat[]{
        \includegraphics[width=0.32\textwidth]{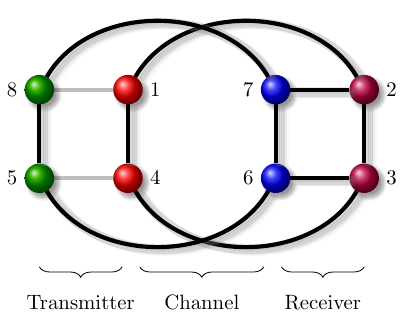}
        \label{subfig:  octagon cyclic6 divided}
    }
    \caption{
        \protect\subref{subfig: octagon cyclic5} and \protect\subref{subfig: octagon cyclic6} are the cluster patterns in the illustrative network. \protect\subref{subfig:  octagon cyclic5 divided} and \protect\subref{subfig:  octagon cyclic6 divided} are the split of the nodes of the network in \protect\subref{subfig: octagon cyclic5} and \protect\subref{subfig: octagon cyclic6} between the transmitter and receiver, respectively.}
    \label{fig: octagons}
\end{figure}

\begin{figure}
    \centering
    \subfloat[]{
        \includegraphics[width=0.75\textwidth]{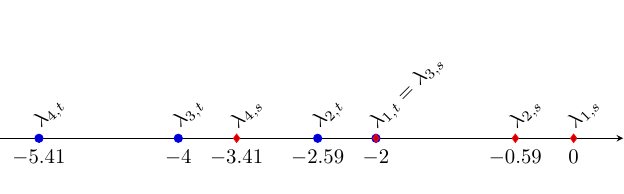}
        \label{subfig: distribution of eigenvalues 1}
    } \\[-1em]
    \subfloat[]{
        \includegraphics[width=0.75\textwidth]{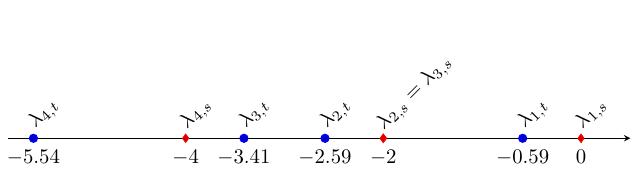}
        \label{subfig: distribution of eigenvalues 2}
    } \\
    \caption[]{
        The distribution of eigenvalues with respect to \protect\subref{subfig: distribution of eigenvalues 1} $\mathcal{S}_{11}$ and $\mathcal{S}_{22}$, and \protect\subref{subfig: distribution of eigenvalues 2} $\mathcal{S}_{12}$ and $\mathcal{S}_{21}$.
    }
\end{figure}

\begin{figure*}
    \centering
    \subfloat[]{
        \includegraphics[width=0.464\textwidth]{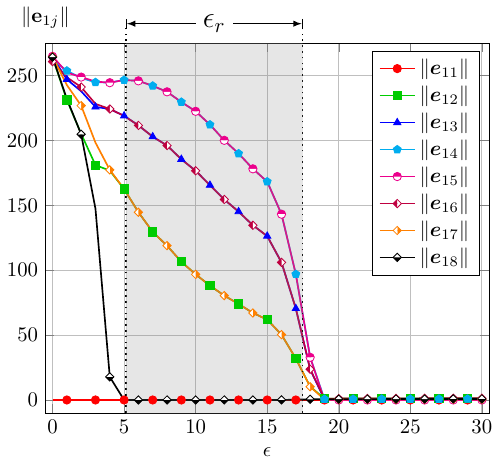}
        \label{subfig: coupling strength change for symbol 0}
    } \hfil
    \subfloat[]{
        \includegraphics[width=0.45\textwidth]{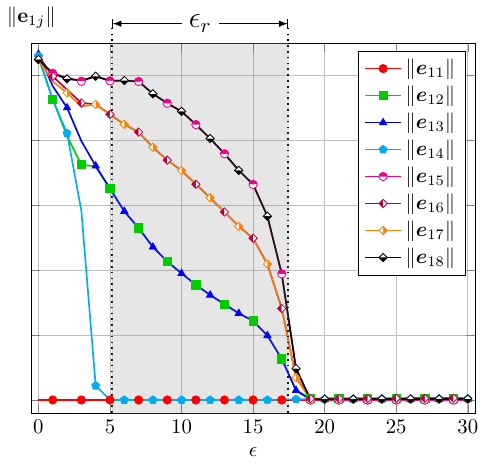}
        \label{subfig: coupling strength change for symbol 1}
    }
    \caption[]{
    \protect\subref{subfig: coupling strength change for symbol 0} The spatiotemporal status of the network when $s_k=s_1=0$ (\figref{subfig: octagon cyclic5 divided}).
    \protect\subref{subfig: coupling strength change for symbol 1} and when $s_k=s_2=1$ (\figref{subfig: octagon cyclic6 divided}) with respect to coupling strength $\epsilon$. The range $\epsilon_r=[5.15, 17.46]$ is the guaranteed cluster synchronization range.
    }
    \label{fig: coupling strength change}
\end{figure*}

Consider the networks in \figref{subfig: octagon cyclic5} and \ref{subfig: octagon cyclic6} consist of the classical Chen systems
\begin{equation}
    \dot{\bm{x}}_i =
    f(\bm{x}_i) =
    \left[\matrix{
            a (x_{i,2} - x_{i,1})                  \cr
            x_{i,1} (c - a - x_{i,3}) +  c x_{i,2} \cr
            x_{i,1} x_{i,2} - b x_{i,3}            \cr
        }\right], \quad
    i = 1, 2 \ldots, 8.
    \label{eq: chen system}
\end{equation}
where $\bm{x}_i = [x_{i,1}, x_{i,2}, x_{i,3}]^T, a = 35, b = 8 / 3, c = 28$ , with inner coupling matrix
\begin{equation}
    \bm{\Gamma} =
    \left[\matrix{
            0 & 0 & 0 \cr
            1 & 0 & 0 \cr
            0 & 0 & 0 \cr
        }\right].
\end{equation}
The networks have the coupling matrices
\begin{equation}
    \bm{\Xi}_1 =
    \left[\matrix{
            -2 & ~1 & ~0 & ~0 & ~0 & ~0 & ~0 & ~1 \cr
            ~1 & -3 & ~1 & ~0 & ~0 & ~0 & ~1 & ~0 \cr
            ~0 & ~1 & -3 & ~1 & ~0 & ~1 & ~0 & ~0 \cr
            ~0 & ~0 & ~1 & -2 & ~1 & ~0 & ~0 & ~0 \cr
            ~0 & ~0 & ~0 & ~1 & -2 & ~1 & ~0 & ~0 \cr
            ~0 & ~0 & ~1 & ~0 & ~1 & -3 & ~1 & ~0 \cr
            ~0 & ~1 & ~0 & ~0 & ~0 & ~1 & -3 & ~1 \cr
            ~1 & ~0 & ~0 & ~0 & ~0 & ~0 & ~1 & -2 \cr
        }\right]
\end{equation}
and
\begin{equation}
    \bm{\Xi}_2 =
    \left[\matrix{
            -2 & ~1 & ~0 & ~1 & ~0 & ~0 & ~0 & ~0 \cr
            ~1 & -3 & ~1 & ~0 & ~0 & ~0 & ~1 & ~0 \cr
            ~0 & ~1 & -3 & ~1 & ~0 & ~1 & ~0 & ~0 \cr
            ~1 & ~0 & ~1 & -2 & ~0 & ~0 & ~0 & ~0 \cr
            ~0 & ~0 & ~0 & ~0 & -2 & ~1 & ~0 & ~1 \cr
            ~0 & ~0 & ~1 & ~0 & ~1 & -3 & ~1 & ~0 \cr
            ~0 & ~1 & ~0 & ~0 & ~0 & ~1 & -3 & ~1 \cr
            ~0 & ~0 & ~0 & ~0 & ~1 & ~0 & ~1 & -2 \cr
        }\right]
\end{equation}
and the cluster patterns $\mathscr{C}_1 = \{\{1, 8\}, \{2, 7\}, \{3, 6\}, \{4, 5\}\}$ and $\mathscr{C}_2 = \{\{1, 4\}, \{2, 3\}, \{5, 8\}, \{6, 7\}\}$ corresponding to the symmetries $\mathcal{S}_1$ and $\mathcal{S}_2$, respectively. The permutation matrices corresponding to $\mathcal{S}_1$ and $\mathcal{S}_2$ are
\begin{equation}
    \bm{\Delta}_1 =
    \left[\matrix{
            ~1 & ~0 & ~0 & ~0 & ~0 & ~0 & ~0 & ~1 \cr
            ~0 & ~1 & ~0 & ~0 & ~0 & ~0 & ~1 & ~0 \cr
            ~0 & ~0 & ~1 & ~0 & ~0 & ~1 & ~0 & ~0 \cr
            ~0 & ~0 & ~0 & ~1 & ~1 & ~0 & ~0 & ~0 \cr
            ~0 & ~0 & ~0 & ~1 & ~1 & ~0 & ~0 & ~0 \cr
            ~0 & ~0 & ~1 & ~0 & ~0 & ~1 & ~0 & ~0 \cr
            ~0 & ~1 & ~0 & ~0 & ~0 & ~0 & ~1 & ~0 \cr
            ~1 & ~0 & ~0 & ~0 & ~0 & ~0 & ~0 & ~1 \cr
        }\right]
\end{equation}
and
\begin{equation}
    \bm{\Delta}_2 =
    \left[\matrix{
            ~1 & ~0 & ~0 & ~1 & ~0 & ~0 & ~0 & ~0 \cr
            ~0 & ~1 & ~1 & ~0 & ~0 & ~0 & ~0 & ~0 \cr
            ~0 & ~1 & ~1 & ~0 & ~0 & ~0 & ~0 & ~0 \cr
            ~1 & ~0 & ~0 & ~1 & ~0 & ~0 & ~0 & ~0 \cr
            ~0 & ~0 & ~0 & ~0 & ~1 & ~0 & ~0 & ~1 \cr
            ~0 & ~0 & ~0 & ~0 & ~0 & ~1 & ~1 & ~0 \cr
            ~0 & ~0 & ~0 & ~0 & ~0 & ~1 & ~1 & ~0 \cr
            ~0 & ~0 & ~0 & ~0 & ~1 & ~0 & ~0 & ~1 \cr
        }\right]
\end{equation}

\begin{figure}
    \centering
    \includegraphics[width=0.45\textwidth]{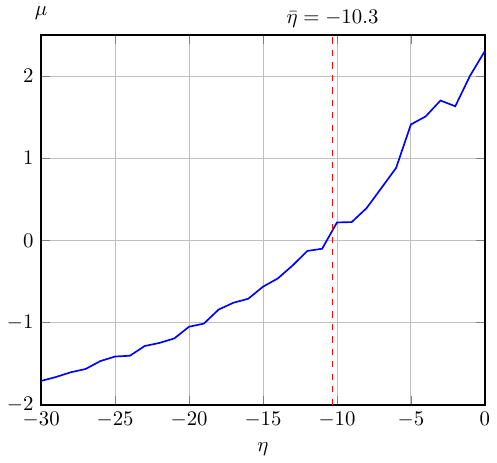}
    \caption[]{The master stability function calculated numerically for \eref{eq: chen system} using \eref{eq: linearized system}. The stability threshold $\bar{\eta} \approx -10.3$ for which $\mu  = g(\eta) < 0, \eta \leq \bar{\eta}$ is indicated by dashed vertical line.}
    \label{subfig-error-versus-epsilon}
\end{figure}

The distribution of the eigenvalues corresponding to $\mathcal{S}_1$ and $\mathcal{S}_2$ is shown in \figref{subfig: distribution of eigenvalues 1}. Since $|\lambda_{2}^s| < \lambda_{min}$, the eigenvalue condition is satisfied for both $\mathcal{S}_{1}$ and $\mathcal{S}_2$. \figref{subfig-error-versus-epsilon} depicts the master stability function $g(\cdot)$ calculated for \eref{eq: chen system} using \eref{eq: linearized system}. We have $\bar{\eta} \approx -10.3$ for \eref{eq: chen system}. Thus, $\epsilon_{r} = [5.15, 17.46]$ is the range for $\epsilon$ to realize both $\mathscr{C}_1$ and  $\mathscr{C}_2$.  \figref{fig: coupling strength change} depicts the change of the spatio-temporal status of the networks in \figref{subfig:  octagon cyclic5 divided} and \ref{subfig: octagon cyclic6}. The nodes are de-synchronized for $\epsilon < \epsilon_l \approx 4.5$, while all the nodes are completely synchronized (full synchronization) for $\epsilon > \epsilon_h \approx 19$.  We have the cluster patterns $\mathscr{C}_1$ and $\mathscr{C}_2$ in the networks \figref{subfig: octagon cyclic5 divided} and \ref{subfig:  octagon cyclic6 divided}, respectively, for $\epsilon \in [\epsilon_l, \epsilon_h] \supset \epsilon_r $,

The network in \figref{subfig: octagon cyclic6} is obtained from the network in \figref{subfig: octagon cyclic5} by removing the links between the node pairs $(1, 8)$ and $(4, 5)$  and connecting the node pairs $(1, 4)$ and $(5, 8)$. Hence, if we choose $\epsilon \in \epsilon_r$, it is possible to switch between $\mathscr{C}_1$ and  $\mathscr{C}_2$ by controlling only the links $\ell_{14}, \ell_{45}, \ell_{58}, \ell_{18}$ without touching the remaining links.

Note that the networks in \figref{subfig: octagon cyclic5} and \figref{subfig: octagon cyclic6} have also $\mathcal{S}_{3}$ and $\mathcal{S}_{4}$ symmetries, respectively. However, from the eigenvalue distribution illustrated in \figref{subfig: distribution of eigenvalues 2}, the eigenvalue condition in (\ref{eq-eigvalue-condition}) is not satisfied. Thus, the cluster patterns corresponding to $\mathcal{S}_{3}$ and $\mathcal{S}_{4}$ cannot be observed in the networks in \figref{subfig: octagon cyclic5} and \figref{subfig: octagon cyclic6}.

\subsubsection{Signalling}
\label{subsec: signalling}

\begin{figure*}
    \centering
    \includegraphics[width=\textwidth]{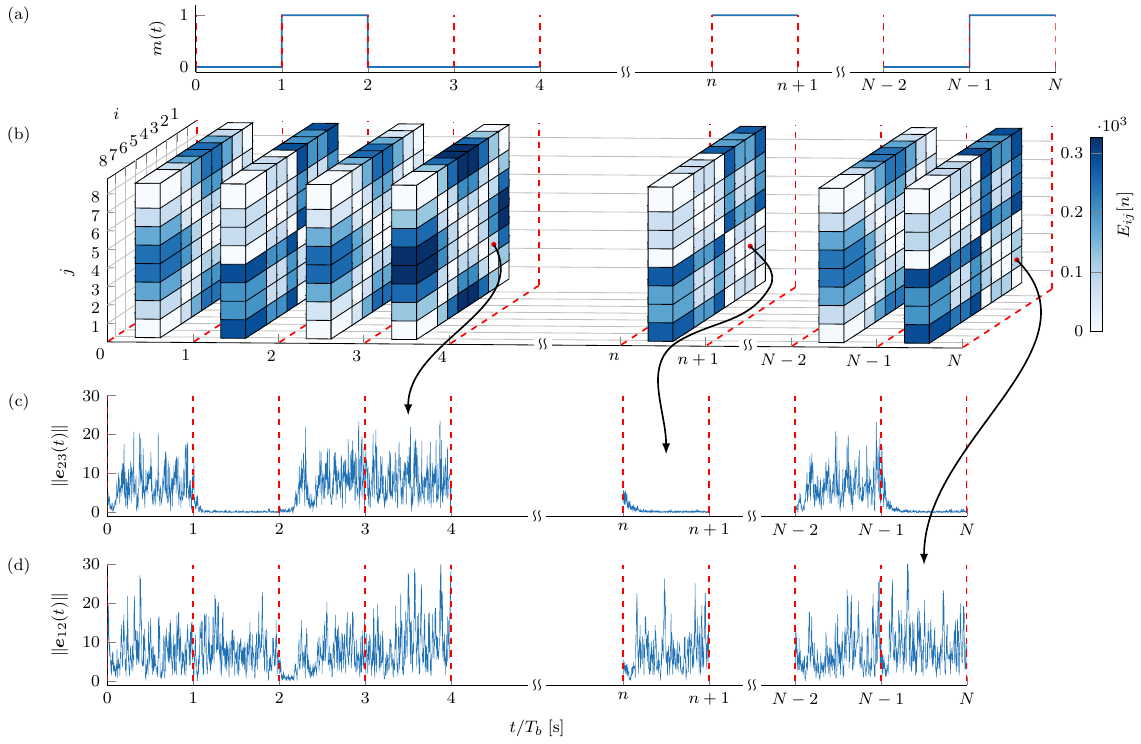}
    \caption[]{Transmission of symbols $s[n] \in \{0, 1\}, n = 0, 2, \ldots, N$. (a) Time waveform of message signal $m(t) = s[n], n T_b \leq t \leq (n + 1) T_b$, where $T_b$ is  the symbol duration. (b) The cluster patterns $\mathscr{C}[n]$. (c) Time waveform, $\Vert \bm{e}_{23}(t) \Vert$, in the receiver. (d) Time waveform, $\Vert \bm{e}_{12}(t) \Vert$, over a channel link.}
    \label{fig: signaling}
\end{figure*}

\begin{figure*}
    \centering
    \includegraphics[width=\textwidth]{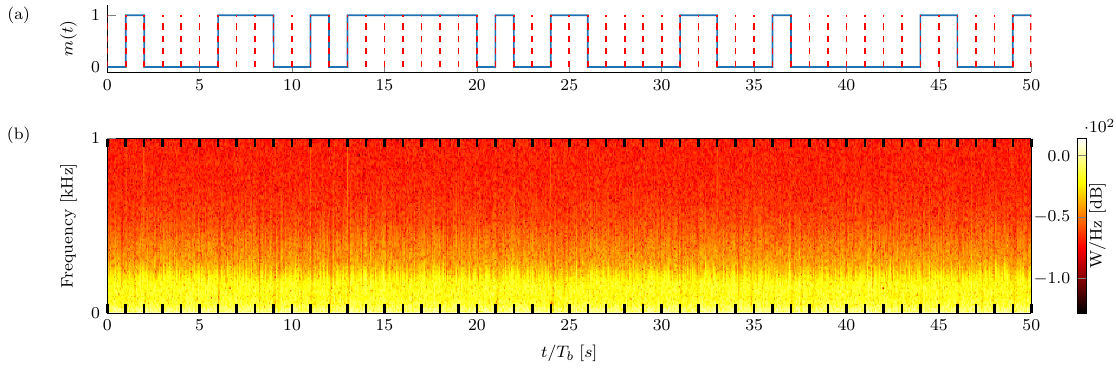}
    \caption[]{ (a) The time waveform of message signal $m(t) = s[n]$, where $T_b$ is  the symbol duration (b) The spectrogram of $\Vert \bm{e}_{12} \Vert$ during the transmission of symbols $s[n] \in \{0, 1\}$.}
    \label{fig: spectrogram}
\end{figure*}

We divide the nodes of the networks in \figref{subfig: octagon cyclic5} and  \figref{subfig: octagon cyclic6} between the transmitter and receiver as shown in \figref{subfig: octagon cyclic5 divided} and \figref{subfig:  octagon cyclic6 divided}, respectively.  As the cluster pattern capacity of the network in \figref{fig: octagons} is $M=2$, we can use the network in a binary communication system in which the symbols $s_1 = 0, s_2 = 1$ are encoded with $\mathscr{C}_1$ and $\mathscr{C}_2$, respectively.

We assume that the channel noise is present only on the links $\ell_{12}, \ell_{34}, \ell_{56}, \ell_{78} \in \mathcal{L}_C$. \figref{fig: signaling} illustrates the signaling of the communication system \eref{eq: chen system} with $\epsilon = 7$ and $\sigma = 0.25$. The error waveforms in \figref{fig: signaling} illustrate $\mathscr{C}_k$ during the transmission of $s_k$ for $k = 1, 2$. Note that by construction, the nodes of the links in the channel do not synchronize with each other regardless of the symbol being sent during the whole transmission.

If the signals flowing through the channel are eavesdropped, the information being transmitted cannot be extracted without the knowledge of the network topologies and symbol-to-cluster-pattern encoding. The spectrogram given in \figref{fig: spectrogram} shows that the frequency content of the transmitted signals is independent of the symbols sent, apparently. The signals transmitted through the channel do not carry information about the transmitted information directly, rather, the information is distributed over the network. This very property of the proposed approach provides the covertness.

\subsubsection{Noise Performance}
\label{subsec: noise performance}

The performance of the proposed scheme is degraded by synchronization errors as well as the noise acting on the channel. Occasionally, the error energy during the bit duration is not small/large enough, although the corresponding receiver nodes tend to synchronize/desynchronize in reality for a larger bit duration because of the asymptotic nature of the synchronization. Therefore, $T_b$  must be chosen large enough to allow the network to settle on the synchronization manifolds for the respective symbols. The second source of detection errors is the diffusive term in \eref{eq:-odenet}. Obviously, the dominant dynamics cannot be considered chaotic anymore and the synchronization manifolds are destroyed completely when the noise amplitude is large enough.

\begin{figure}
    \centering
    \includegraphics[width=0.5\textwidth]{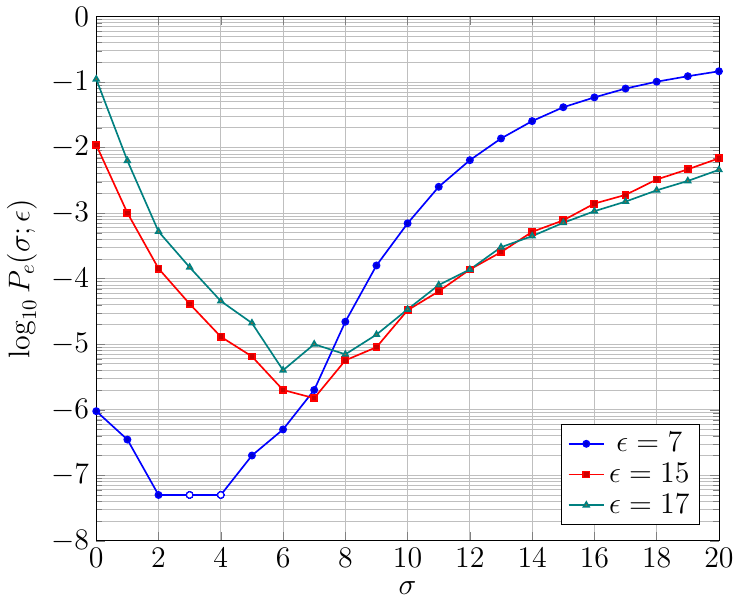}
    \caption[]{BER performance in terms of noise level $\sigma$ and coupling strength $\epsilon$ in \eref{eq: sdenet}. The points marked with \protect\tikz\protect\draw[blue, fill=white, thick] (0,0) circle (2pt); presents the cases with no bit errors, for which the worst case is assumed in calculation of $P_e$.
    }
    \label{fig: ber in terms of snr}
\end{figure}

We solved \eref{eq: sdenet} using Order 1.5 strong Stochastic Runge-Kutta (SRK) method with weak order 2 to demonstrate the noise performance\cite{rossler2010runge}. The simulation results have been obtained for \eref{eq: chen system} and a spreading factor (the number of samples in one symbol duration) of $200$. The transmitted bits have been chosen randomly with equal probabilities. The number of bits transmitted has been adapted to the noise levels as in Table \ref{tbl: adaptation of bits}. We assumed reliable transmission is possible when the probability of making an incorrect decision at the receiver given the noise level $\sigma$ and coupling strength $\epsilon \in \epsilon_r$ in \eref{eq: sdenet}, i.e., $P_e(\sigma;\epsilon)$, is smaller than $10^{-6}$. \figref{fig: ber in terms of snr} depicts the bit error rate (BER) performances of the system in the worst case. Clearly, the BER performance depends on the coupling strength $\epsilon$.

$P_e(\sigma; \epsilon=7) < P_e(\sigma; \epsilon=15) < P_e(\sigma; \epsilon=17) $ for $0<\sigma<7$. In comparison, the network undergoes complete desynchronization to cluster synchronization followed by full synchronization for increasing values of $\epsilon$ as suggested by \figref{fig: coupling strength change}. The distribution of the synchronization error energies between the node pairs does not change uniformly with $\epsilon$. The average synchronization error energies between the node pairs converge as $\epsilon$ approaches its lower and upper limits. The network is on the verge of complete desynchronization and complete synchronization for $\epsilon=5$ and $\epsilon=17$ respectively. The detection of the correct cluster pattern is more unlikely for both cases. We get lower detection error probabilities if the synchronization error energies between the intra-cluster nodes are more dispersed for both symbols. In \figref{fig: octagons}, we can see that the difference between cluster to non-cluster error energies are maximum around $\varepsilon\approx 7$ for both symbols (\figref{fig: coupling strength change}) at which BER performance is better (\figref{fig: ber in terms of snr}).

Basically, we have two trends in the BER performance. The asymptotic nature of the stability of synchronization manifolds causes bit errors when switching from one manifold to the other. The bit errors occur when the time required for synchronization is longer than the selected bit duration. For the low noise levels ($\sigma <2$), the errors due to manifold switching are dominant, and the network needs more time to form the required cluster synchronization patterns for obtaining lower error probabilities. As we increase the noise to moderate levels ($2<\sigma<5$), the presence of the noise helps the desynchronization of the nodes that are not synchronized in the newly switched pattern, yielding better BER performance compared to zero noise level. No bit errors observed for $\epsilon=7$ and $\sigma=3$ or $\sigma=4$ in the numerical simulations. Hence, $P_e(\sigma;\epsilon=7) \leq 5 \cdot 10^{-8}$ can be assumed safely in this range $\sigma \in (3, 4)$. For higher noise levels ($\sigma>7$), the dynamics of the system dominated by the diffusion term in \eref{eq: sdenet} cannot be  considered chaotic anymore. The channel noise destroys the cluster patterns eventually, and becomes the primary cause of the detection errors.

\begin{table}
    \centering
    \caption{Number of bits used for different noise levels in \figref{fig: ber in terms of snr}.}
    \label{tbl: adaptation of bits}
    \begin{tabular}{|c|cccc|}
        \hline
        $\sigma$ range & $[0, 4]$       & $[5, 10]$      & $[11, 15]$     & $[16, 20]$     \\  \hline
        Number of bits & $2 \cdot 10^7$ & $1 \cdot 10^6$ & $5 \cdot 10^5$ & $1 \cdot 10^5$ \\
        \hline
    \end{tabular}
\end{table}

\subsection{Example II: Design with Control Network}
\label{subsec-an-example-network-with-control}

\subsubsection{The Network}

\begin{figure}
    \centering
    \subfloat[]{
        \includegraphics[width=0.4\textwidth]{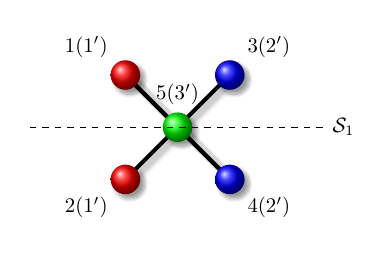}
        \label{subfig-trial-2-pattern-1}
    } \hfil
    \subfloat[]{
        \includegraphics[width=0.4\textwidth]{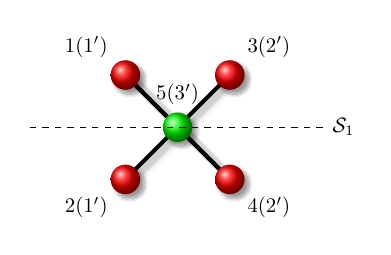}
        \label{subfig-trial-2-pattern-2}
    }
    \caption[]{The cluster patterns \protect\subref{subfig-trial-2-pattern-1} $\mathscr{C}_1$ and \protect\subref{subfig-trial-2-pattern-2} $\mathscr{C}_2$}
    \label{fig-cluster-patterns-example2}
\end{figure}

\begin{figure}
    \centering
    \subfloat[]{
        \includegraphics[width=0.315\textwidth]{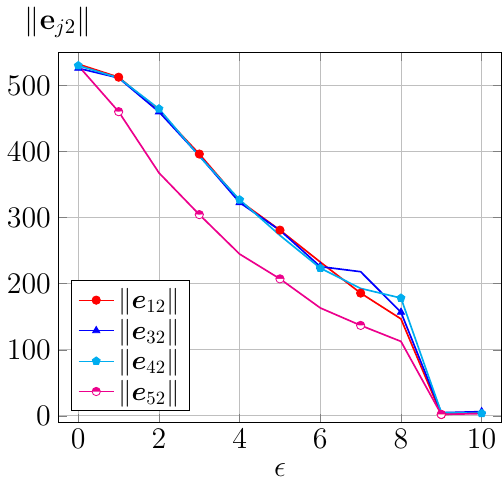}
        \label{subfig-pinning-control-error-versus-epsilon}
    }
    \subfloat[]{
        \includegraphics[width=0.315\textwidth]{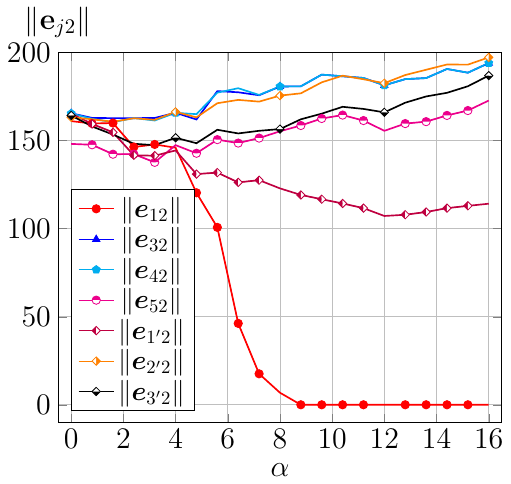}
        \label{subfig-pinning-control-error-versus-alpha}
    } 
    \subfloat[]{
        \includegraphics[width=0.315\textwidth]{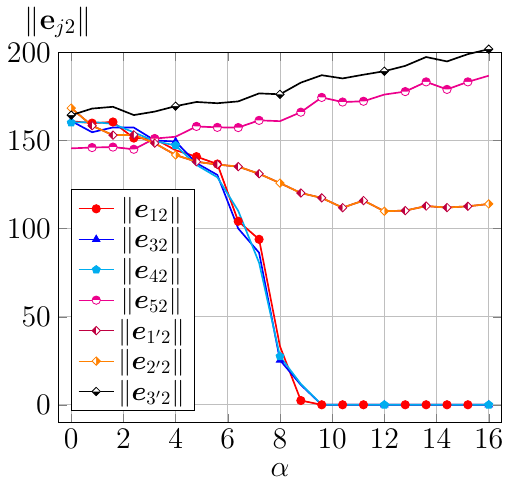}
        \label{subfig-pinning-control-error-versus-alpha2}
    }
    \caption{The change of spatiotemporal status of the network with respect to \protect\subref{subfig-pinning-control-error-versus-epsilon} coupling strength $\epsilon$, and \protect\subref{subfig-pinning-control-error-versus-alpha} and \protect\subref{subfig-pinning-control-error-versus-alpha2} control strength $\alpha$ corresponding to the cluster patterns in \ref{subfig-trial-2-pattern-1} and \ref{subfig-trial-2-pattern-2}, respectively.}
\end{figure}

Consider the network and desired cluster patterns $\mathscr{C}_1$ and $\mathscr{C}_2$ illustrated in \figref{fig-cluster-patterns-example2}. The control node of each node has been written in the parentheses next to it. \figref{subfig-pinning-control-error-versus-epsilon} illustrates the change of the spatiotemporal status of the network without the control. Note that the network undergoes from complete de-synchronization to full synchronization directly, without achieving any cluster synchronization pattern. Hence, to achive the cluster patterns $\mathscr{C}_1$ and $\mathscr{C}_2$ in \figref{fig-cluster-patterns-example2}, we have to apply control the network. For the network in \figref{subfig-trial-2-pattern-1}, we have the coupling and control matrix
\begin{equation}
    \bm{\Xi} = \left[
        \matrix{
            -1  & ~~0 & ~~0 & ~~0 & ~~~1 \cr
            ~~0 & -1  & ~~0 & ~~0 & ~~~1 \cr
            ~~0 & ~~0 & -1  & ~~0 & ~~~1 \cr
            ~~0 & ~~0 & ~~0 & -1  & ~~~1 \cr
            ~~1 & ~~1 & ~~1 & ~~1 & -4 \cr
        } \right],
        \quad 
        \bm{A} = \left[
        \matrix{
            -1  & ~~0 & ~~~1 \cr
            ~~0 & -1  & ~~~1 \cr
            ~~2 & ~~2 & -4 \cr
        }
        \right],
\end{equation}
respectively, and the interconnection matrix
\begin{equation}
    \bm{\Delta} = \left[
        \matrix{
            1 & 0 & 0 \cr
            1 & 0 & 0 \cr
            0 & 1 & 0 \cr
            0 & 1 & 0 \cr
            0 & 0 & 1 \cr
        }
        \right]
\end{equation}
that connects the control and controlled network. The change of spatiotemporal status of the network with respect to the change of the control strength $\alpha$ is shown in \figref{subfig-pinning-control-error-versus-alpha}. It is seen that the cluster pattern shown in \figref{subfig-trial-2-pattern-1} is achieved for $\alpha > 8$, and the control nodes $1^\prime$, $2^\prime$ and $3^\prime$ do not synchronize with any of the nodes in their corresponding clusters.

The results in \figref{subfig-pinning-control-error-versus-alpha} shows that the cluster pattern in \figref{subfig-trial-2-pattern-2} cannot be achieved without applying a control network. For the cluster pattern $\mathscr{C}_2$ in \figref{subfig-pinning-control-error-versus-alpha2}, we want the nodes $1, 2, 3, 4$ to be synchronized. Since the control nodes $1^\prime$ and $2^\prime$ control the node pairs $(1, 2)$ and $(3, 4)$, respectively, if the control nodes $1^\prime$ and $2^\prime$ synchronize to each other, then we can achieve the cluster pattern shown in \ref{subfig-trial-2-pattern-2}. The nodes $1^\prime$ and $2^\prime$ synchronize for the control matrix
\begin{equation}
    \bm{A} = \left[
        \matrix{
            -20  & ~~20 & 0 \cr
            ~~20 & -20  & 0 \cr
            ~~~0 & ~~~0 & 0 \cr
        } \right].
\end{equation}

The change of the spatiotemporal status of the network with respect to the changing values of the control strength $\alpha$ is given in \figref{subfig-pinning-control-error-versus-alpha2}. Note from \figref{subfig-pinning-control-error-versus-alpha2} that the cluster pattern shown in \figref{subfig-trial-2-pattern-2} is achieved for $\alpha > 10$. As we have achieved two different cluster patterns, we can encode the symbols $0$ and $1$ of the communication system by these cluster patterns.

\begin{figure}
    \centering
    \subfloat[]{
        \includegraphics[width=0.475\textwidth]{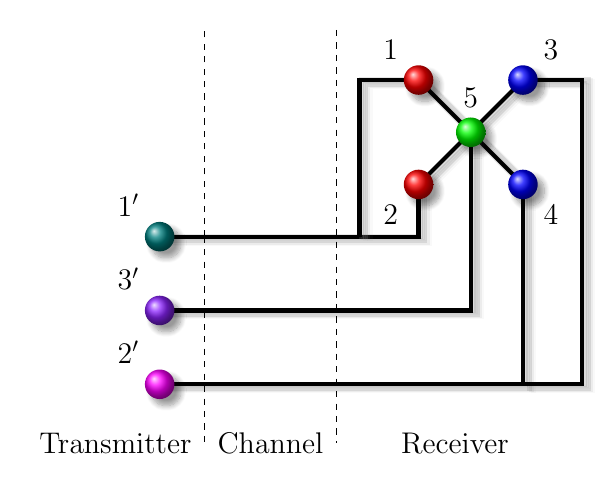}
        \label{subfig-symbol0-transmission}
    } \hfil
    \subfloat[]{
        \includegraphics[width=0.475\textwidth]{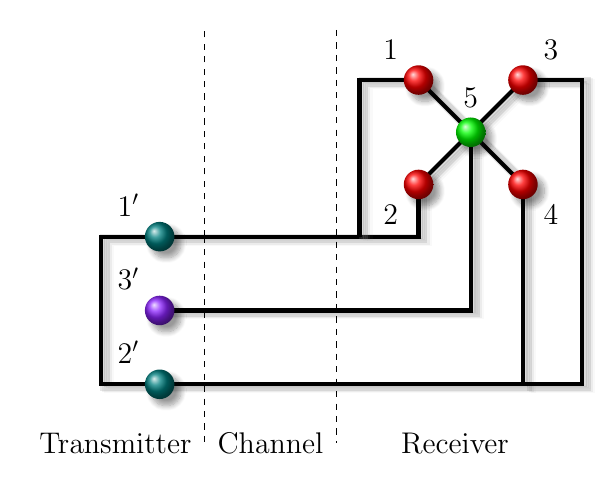}
        \label{subfig-symbol1-transmission}
    }
    \caption{\ref{sub@subfig-symbol0-transmission} and \ref{sub@subfig-symbol1-transmission} The split of nodes of the network in \figref{subfig-trial-2-pattern-1} and \figref{subfig-trial-2-pattern-2}, respectively.}
    \label{fig-split-of-nodes}
\end{figure}

We can split the nodes as shown in \figref{subfig-symbol0-transmission} and in \figref{subfig-symbol1-transmission}, and encode the symbols $0$ and $1$ to the cluster patterns in \figref{subfig-symbol0-transmission} and \figref{subfig-symbol1-transmission}, respectively. The node $1^\prime$ is coupled to node $2^\prime$ during symbol $0$ transmission, and we decouple $1^\prime$ and $2^\prime$ during symbol $1$ transmission. The nodes are split such that the control nodes stay at the transmitter side while the nodes of the original network stay at the receiver side. In this way, we can control the cluster pattern of the whole network from the transmitter. The control nodes do not synchronize to the nodes in the receiver during the whole transmission. Since the coupling between the control and controlled nodes are unidirectional and the nodes are always in their chaotic regimes, the signals transmitted through the channel are always chaotic.

\subsubsection{Noise Performance}

\begin{table}
    \centering
    \caption{Number of bits used for different values of $\sigma, \alpha$ and $s_f$ in \figref{fig-ber2-in-terms-of-snr}.}
    \label{tbl: adaptation of bits2}
    \small 
    \begin{tabular}{|c|c|c|c|}
        \hline 
        $\alpha$ & $s_f$ & $\sigma$ range & Number of bits\\ \hline  
        \multirow{3}{*}{20} & \multirow{3}{*}{500} & [0, 0.15] & $1 \cdot 10^7$ \\ 
         &  & [0.2, 0.65] & $5 \cdot 10^6$ \\ 
         &  & [0.7, 1.0] & $1 \cdot 10^5$ \\ \hline  
        \multirow{2}{*}{20} & \multirow{2}{*}{1000} & [0, 0.15] & $1 \cdot 10^7$ \\ 
         &  & [0.2, 0.65] & $7 \cdot 10^6$ \\  
         &  & [0.7, 1.0] & $1 \cdot 10^6$ \\ \hline  
        \multirow{4}{*}{10} & \multirow{4}{*}{1000} & [0, 0.6] & $1 \cdot 10^7$ \\ 
         &  & [0.65, 1] & $1 \cdot 10^6$ \\ 
         &  & [1, 1.5] & $1 \cdot 10^5$ \\ \hline  
    \end{tabular}
\end{table}

\begin{figure}
    \centering
    \subfloat[]{
        \includegraphics[width=0.48\textwidth]{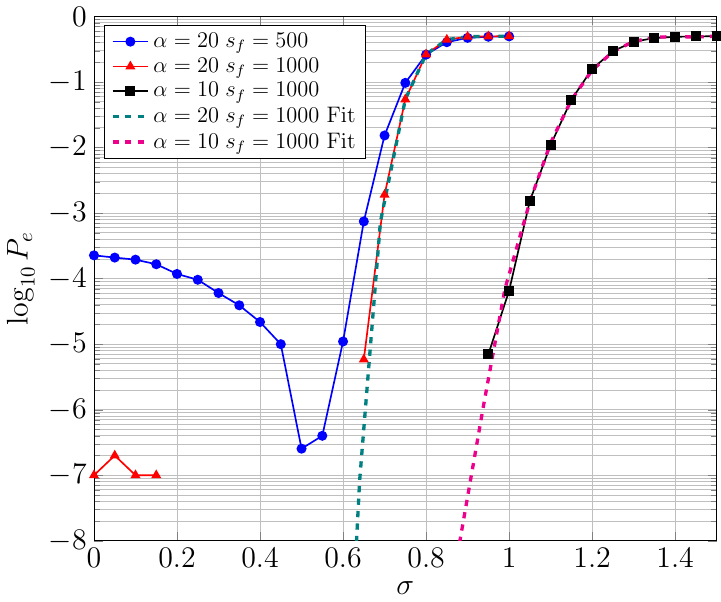}
        \label{fig-ber2-in-terms-of-snr}
    }
    \subfloat[]{
        \includegraphics[width=0.48\textwidth]{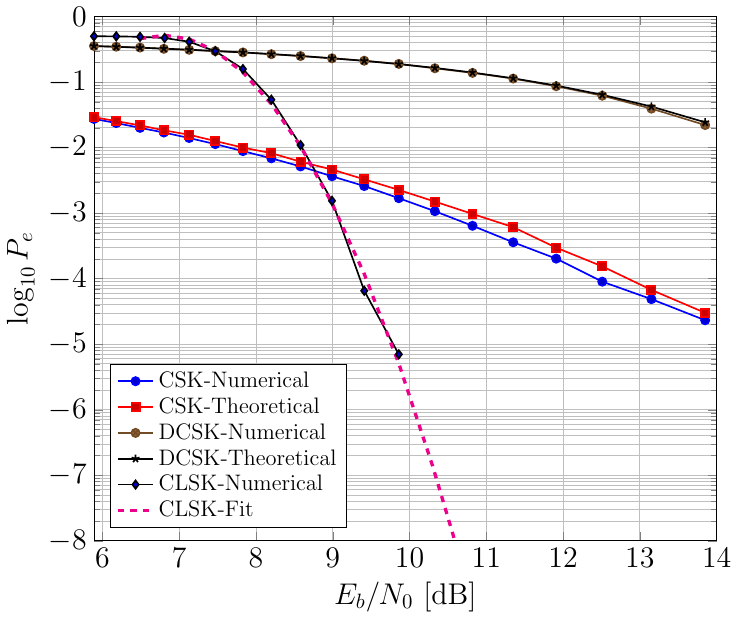}
        \label{fig-ber-csk-dcsk}
    }
    \caption{
        BER performances (plotted by solid lines with markers) obtained by numerical simulations and curves (plotted by dashed lines) fitted to the BER performances to represent the trends of the noise performances.
        \protect\subref{fig-ber2-in-terms-of-snr} BER performances in terms of noise level $\sigma$ and probability of making a detection error $P_e$ of the network in \figref{fig-split-of-nodes} for $\epsilon = 3$ and different values of $\alpha$ and $s_f$.
        \protect\subref{fig-ber-csk-dcsk} BER performances of a single user CSK, DCSK, and CLSK scheme for $\epsilon=3, \alpha=10, s_f=1000$ in terms of the ratio of average symbol energy to noise power spectral density $E_b/N_0$ versus the probability of making a detection error $P_e$.
    }
\end{figure}

The noise performance of the network in \figref{fig-split-of-nodes} is given in \figref{fig-ber2-in-terms-of-snr} in terms of $\sigma$ in (\ref{eq: sdenet compact}) versus the probability of making a detection error $P_e$. Table \ref{tbl: adaptation of bits2} shows the adaptation of the number of bits used in the simulation in \figref{fig-ber2-in-terms-of-snr} for different values of $\alpha, \sigma$, and $s_f$ defined as the number of samples in the transmitted signals within a symbol duration. Similar to the previous example in \S\ref{subsec-an-example-network-without-control}, we have two different trends in the noise performance. For a fixed network where all its parameters, i.e., coupling and control strenghts, are kept constant, the ambient noise is the primary source of symbol detection error. Once the noise strength decreases and passes through a threshold that is dependent on the network parameters, the numerical simulation errors are the main cause of the symbol detection errors. 

The cluster synchronization defined in (\ref{eq-cluster-synchronization-definition}) assumes asymptotical synchronization, where the synchronization errors between the nodes in the same clusters decay to zero as the time approaches infinity. However, practically, we have to specify a certain finite symbol duration. During the symbol transmission, the nodes in the same cluster may need more time to be decided as synchronized. The detection decision given in the specified symbol duration results in detection errors. This type of numerical simulation error can be overcome using finite-time synchronization techniques in which the nodes are guaranteed to be synchronized within a specified amount of time. This observation is justified numerically in  \figref{fig-ber2-in-terms-of-snr}. Doubling the symbol duration, i.e., doubling the spreading factor $s_f$, lowers the probability of detection errors for the same amount of noise. 

The formation and deformation of the cluster patterns of the network are controlled by the control strength $\alpha$. It is the networks' not being able to deform a synchronized cluster pattern that causes the numerical simulation errors. Hence, for excessive values of control strength, it becomes harder the deform a synchornized cluster pattern. In \figref{fig-ber2-in-terms-of-snr}, is shown the effect of the control strength on the noise performance. Lowering the control strength from $\alpha = 20$ to $\alpha=10$ improves the noise performance dramatically and removes the numerical simulation errors for excessively low noise levels. 

\begin{figure}
    \includegraphics[width=\textwidth]{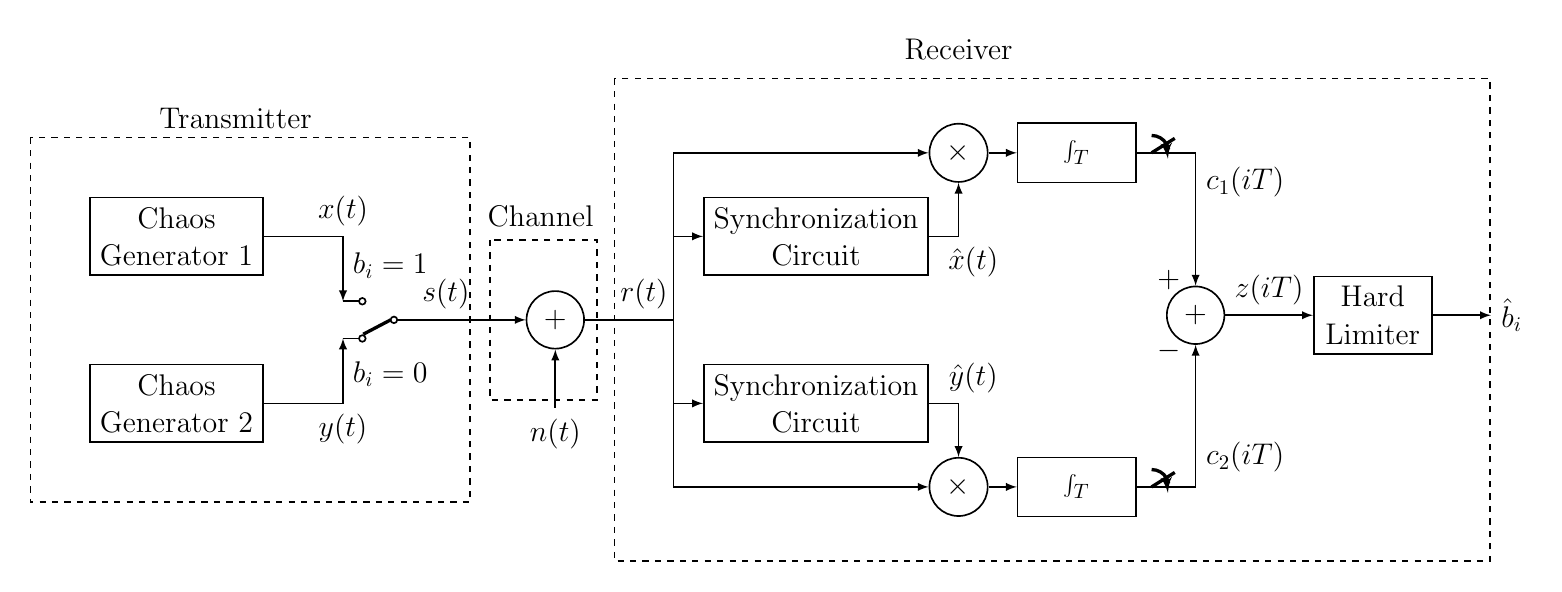}
    \caption{Block diagram of a CSK system. $x(t)$ and $y(t)$ are the chaotic signals and $b_i \in \{0, 1\}, i = 1, 2, \ldots$ are the symbols.$n(t)$ is additive white Gaussian noise. $x(t)$ and $\hat{y}(t)$ are exacta replicas of $x(t)$ and $y(t)$, respectively. $T$ is the symbol duration and $\hat{b}_i$ is the detected symbol.} 
    \label{fig-csk-block-diagram}
\end{figure}

In \ref{fig-csk-block-diagram} is shown the block diagram of a chaos shift keying (CSK) scheme. The chaoic signals $x(t)$ and $y(t)$ are switched to form the transmittes signal $s(t)$ with respect to the symbol stream $b_i \in \{0, 1\}, i = 1, 2, \ldots$ for a symbol duration $T$. The additive white Gausian distributed noise $n(t)$ is added to $s(t)$ resulting in $r(t) = s(t) + n(t)$. $r(t)$ is correalated by $x(t)$ and $\hat{y}(t)$, which are \emph{assumed} to be the \emph{exact} replicas of $x(t)$ and $y(t)$, respectively, relying on the drive-response syncrhonization phenomeonon. The symbol $\hat{b}_i$ that corresponds to the maksimum correlation is detected as the transmitted symbol. In \figref{fig-ber-csk-dcsk} is shown the noise performances of a CSK system obtained numerically and theoretically \cite{tse2003chaos} of single-user CSK scheme in terms of probability of making a detection error $P_e$ versus average symbol energy to noise power spectral density $Eb/N0$. The numerical simulation is performed by solving numerically the ordinary differential equations that results in chaotic signals $x(t)$ and $y(t)$. Then, $r(t)$, which is obtained by \emph{directly adding} $n(t)$ to $s(t)$, is correlated by $x(t)$ and $y(t)$ to detect $\hat{b}_i$. Note from \figref{fig-ber-csk-dcsk} that the theoretical and numerical results are in good agreement. However, this approach does not reflect the phsyical reality that $x(t)$ and $y(t)$ are injeted into the dynamics of the synchronization circuits to generate the replicas $x(t)$ and $\hat{y}(t)$. Hence, the dynamics of the systems must be represented by stochastic differential equations (SDEs), instead of ODEs. We have seen in our simulations that in case of SDEs are used to model the system dynamics, the driven chaotic systems in receiver go out of their chaotic regimes and become unstable, even for small values of $\sigma \approx 0.0083$. Hence, a CSK system is not robust under noise physically. \figref{fig-ber-csk-dcsk} compares the performance of the CLSK obtained by solving numerically the SDE in \ref{eq: sdenet compact}, to those of a single-user CSK and differential chaos shift keying (DCSK). The curve $P_e(x) = 10^{ax^2 + b^x + c}$, where $a=-0.5354, b = 7.2835, c = -25.05$, plotted by a dashed line in \figref{fig-ber-csk-dcsk}, can be fit to the numerical noise performance data of the CLSK. Even under unfair comparison, although the performance of the CLSK lags behind that of CSK for high noise levels, the CLSK outperforms DCSK and CSK for noise levels greater than $E_b/N_0 \approx 7.5$ dB and  $E_b/N_0 \approx 8.75$ dB, respectively. Note in particular the drastical differences in the performances of the CSK, DCSK and CLSK as $E_b/N_0$ increases.

\section{Conclusion}
\label{sec: conclusion}

In this paper, we have proposed a new covert communication scheme based on cluster synchronization in networks of chaotic oscillators. The symbols are encoded to the spatio-temporal state of the network as synchronous cluster patterns. We have illustrated how existing cluster synchronization design techniques could construct the network to achieve different clusters for different values of control parameters. Notably, we ensure that the transmitted signals are always chaotic by construction, regardless of the transmitted symbol. The information is distributed over the network, in contrast to its counterparts in which the information is modulated onto the transmitted signals directly. Therefore, they do not carry information directly. The transmitted information cannot be extracted without knowing which symbols are encoded into which cluster patterns, even if the channel has eavesdropped.

Unlike the available communication systems in which the modulation is in the time-frequency domain, the proposed method performs the modulation in spatio-temporal domain, where the neighborhood of the nodes in the network determines the detected symbol spatially.

Conventionally, the noise in the channel is assumed to be additive, and the signal flow is from the transmitter to the receiver. In the proposed approach, we consider that the noise directly affects the systems' dynamics at receiver subnetworks. The injection of noise directly into the dynamics of the receiver subnetworks causes the noise to be non-additive due to the diffusion terms in the stochastic model. Consequently, an analytical approach to determining noise performance is not an easy task. Nevertheless, we have illustrated the typical performances of the specifically-designed networks of chaotic systems and obtained the BERs of the proposed scheme via extensive numerical simulations assuming binary signaling. The results indicate that the system is robust to noise at the channel, and outperforms some well-kwown chaos communication schemes for $E_b/N_0$ values greater than certain thresholds.
 
One important direction of further research is to determine the symbol capacity of large networks with many symmetries to allow M-ary communication with large $M$. The performance also depends on channel filtering, transmission delay, and jamming. These effects require the basic stochastic model to be extended beyond the presented.

The proposed cluster pattern shift keying scheme offers a pertinent way for covert signaling. The scheme can be combined with existing cryptographic techniques and error-correcting codes to provide an additional layer of protection against eavesdropping on a physical level.

\ack
The numerical calculations reported in this paper were performed at TUBITAK ULAKBIM, High Performance and Grid Computing Center (TRUBA resources).

\section*{References}
\bibliographystyle{vancouver}

\end{document}